\documentclass[12pt]{article}
\usepackage{xcolor}
\usepackage{amsmath}
\usepackage[colorlinks=true,linkcolor=blue, citecolor=blue]{hyperref}%
\usepackage{amsfonts}
\usepackage{amssymb}
\usepackage{amsthm}
\usepackage{epsfig}
\usepackage{graphicx}
\usepackage[ruled,vlined]{algorithm2e}
\usepackage{hyperref}
\usepackage[all]{hypcap}
\usepackage{subcaption}
\usepackage{graphicx}
\usepackage[countmax]{subfloat}
\usepackage{setspace}
\usepackage{gensymb}
\usepackage{color}
\usepackage{bm}
\usepackage{authblk}
\usepackage{fullpage}
\usepackage{url}
\usepackage[authoryear]{natbib}
\usepackage{etoolbox}
\usepackage{amsthm}
\usepackage{multirow}

\theoremstyle{remark}

\makeatletter

\newif\ifabbreviation
\pretocmd{\thebibliography}{\abbreviationfalse}{}{}
\AtBeginDocument{\abbreviationtrue}

\usepackage{titlesec}
\titleformat*{\section}{\large\bfseries}
\titleformat*{\subsection}{\normalsize\bfseries}
\titleformat*{\subsubsection}{\normalsize\bfseries}

\titlespacing*{\section}
{0pt}{4pt}{1pt}
\titlespacing*{\subsection}
{0pt}{4pt}{1pt}

\begin{document}

\title{\textbf{Boost-S}: Gradient Boosted Trees for Spatial Data and Its Application to FDG-PET Imaging Data}


\author[1]{Reza Iranzad} 
\author[1]{Xiao Liu} 
\author[1]{W. Art Chaovalitwongse}
\author[2]{Daniel S. Hippe}
\author[3]{Shouyi Wang}
\author[3]{Jie Han}
\author[1]{Phawis Thammasorn}
\author[4]{Chunyan Duan}
\author[5]{Jing Zeng}
\author[2, 5]{Stephen R. Bowen}
\affil[1]{Department of Industrial Engineering, University of Arkansas}
\affil[2]{Department of Radiology, University of Washington}
\affil[3]{Department of Industrial, Manufacturing \& Systems Engineering, University of Texas at Arlington}
\affil[4]{Department of Mechanical Engineering, Tongji University}
\affil[5]{Department of Radiation Oncology, University of Washington}
\date{\vspace{-26pt} }
\maketitle

\begin{abstract}
Boosting Trees are one of the most successful statistical learning approaches that involve sequentially growing an ensemble of simple regression trees (i.e., ``weak learners''). This paper proposes a new gradient \textbf{Boost}ed Trees algorithm for \textbf{S}patial Data (Boost-S) for spatially correlated data with covariate information.  Boost-S integrates the spatial correlation structure into the classical framework of gradient boosted trees. Each tree is grown by solving a regularized optimization problem, where the objective function involves two penalty terms on tree complexity and takes into account the underlying spatial correlation. A computationally-efficient algorithm is proposed to obtain the ensemble trees. The proposed Boost-S is applied to the spatially-correlated FDG-PET (fluorodeoxyglucose-positron emission tomography) imaging data collected during cancer chemoradiotherapy. Our numerical investigations successfully demonstrate the advantages of the proposed Boost-S over existing approaches for this particular application. 
\end{abstract}

\clearpage
\onehalfspacing
\section{Introduction} \label{sec:introduction}
Spatial data refer to an important type of data which arise in a spatial area and are often correlated in space. Capturing such correlation is the centerpiece of statistical analysis of spatial data \citep{cressie2011statistics, Schabenberger2005}. 
Applications of statistical spatial data modeling can be found in a spectrum of scientific and engineering applications ranging from energy \citep{Ezzat2019}, reliability \citep{Liu2018bp, Fang2019}, quality and manufacturing \citep{Zang2019, Wang2016, Yue2020}, environmental and natural process \citep{Guinness2013, Liu2018p, Liu2020}, medical informatics \citep{Yao2017, Yan2019}, etc. 

\vspace{8pt}
\subsection{The Problem Statement}
In this paper, we are concerned with the modeling problem:
\begin{equation}
    Y(\bm{s}) = Z(\bm{x}_{\bm{s}}) + \varepsilon(\bm{s}), \quad\quad \bm{s}\in \mathcal{S}
    \label{eq:model}
\end{equation}
\noindent where $Y(\bm{s})$ represents the observation at a spatial location $\bm{s} \in \mathcal{S}$, $\bm{x}_{\bm{s}}$ is a vector that collects the available features at $\bm{s}$,  $Z(\bm{x})\equiv \mathbb{E}(Y; \bm{x})$ is the mean-value function given $\bm{x}$, and $\varepsilon(\bm{s})$ is an isotropic and weakly stationary spatial noise process with zero-mean and a covariance $C(h)$, where $C(h)=\text{cov}(Y(\bm{s}),Y(\bm{s}'))$ with $h$ being some distance measure between $\bm{s}$ and $\bm{s}'$; for example, the Euclidean distance. 

The goal of this paper is to tackle the modeling problem (\ref{eq:model}) by devising a new \textit{additive-tree-based} method that approximates $Z(\bm{x})$ as follows:
\begin{equation} \label{eq:Z}
Z(\bm{x}) \approx \phi(\bm{x}) = \sum_{k=1}^{K} f_k(\bm{x}), \quad\quad f \in \mathcal{F}
\end{equation}
where $\{f_k(\bm{x})\}_{k=1}^{K}$ represents an ensemble of binary decision trees, and $\mathcal{F}$ represents the tree space. 

Apparently, the problem hinges on \textbf{how the ensemble of trees $\{f_k(\bm{x})\}_{k=1}^{K}$ can be grown, while taking into account the important spatial correlation of $Y(\bm{s})$}. In particular, we exploit the idea of gradient boosting which involves sequentially growing an ensemble of simple regression trees. Mathematically, each tree is added to the ensemble by solving an optimization problem with a carefully chosen objective function. This goal boils down to three fundamental tasks to be addressed in this paper: (\textit{\textbf{i}}) formulate a (regularized) optimization problem that balances the complexity of individual trees and takes into account the spatial correlation associated with $Y(\bm{s})$; (\textit{\textbf{ii}}) devise an algorithm that solves the regularized optimization problem in a computationally efficient manner (so that trees can be added sequentially to the ensemble); and (\textit{\textbf{iii}}) validate the performance of the proposed method on real datasets. 

\vspace{8pt}
\subsection{A Motivating Application} \label{sec:motivating}
\begin{figure}[h!]
	\centerline{\includegraphics[width=0.8\linewidth]{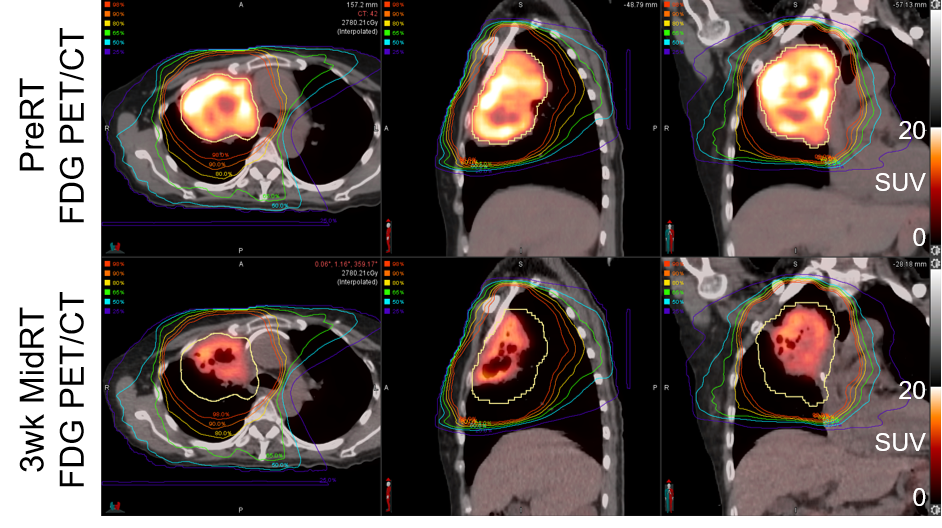}}
	\caption{An illustrative example of spatially-correlated Pre-RT and Mid-RT FDG-PET images}
	\label{fig:SUV}
\end{figure}

A motivating application is first presented. Fluorodeoxyglucose-Positron Emission Tomography (FDG-PET) has been widely used in cancer diagnosis and treatment to detect metabolically active malignant lesions, and plays a critical role in quantitatively assessing and monitoring tumor response to treatment. For illustrative purposes, Figure \ref{fig:SUV} shows the Standardized Uptake Values (SUV) obtained from the FDG-PET imaging of a patient. In this figure, the top row shows the baseline image taken before the radiotherapy (i.e., Pre-RT), while the bottom row shows the FDG PET/CT imaging 3 weeks after radiotherapy (i.e., Mid-RT image). In this case, it is possible to observe the shrinkage of the tumor, indicating the effectiveness of treatment. Hence, the difference between the Mid-RT and Pre-RT images can be naturally used to quantify the tumor's spatial response to treatment. Figure \ref{fig:SUV} also suggests that a tumor typically presents spatially-correlated and spatially-heterogeneous responses. Some areas of a tumor may respond well to treatment while some areas appear to be less responsive. The importance of capturing such spatially-varying and spatially-correlated responses has been discussed in \cite{bowen2019voxel}. 

\begin{figure}[h!]
\vspace{-2mm}
\centerline{\includegraphics[width=\linewidth]{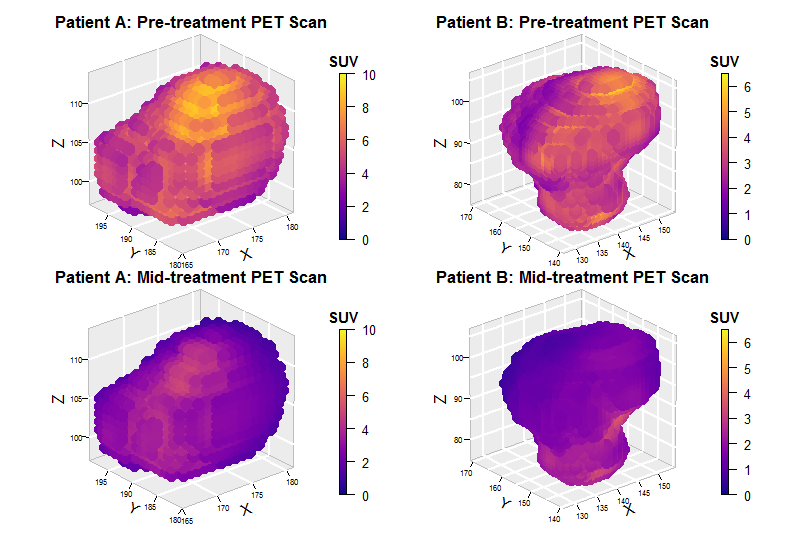}}
\vspace{-3mm}
\caption{3D Pre-RT and Mid-RT SUV images for two patients}
\label{tumors}
\end{figure}

A tumor is a three-dimensional object. 
Figure \ref{tumors} shows the SUV in three-dimensional spaces for another two patients. The top row shows the Pre-RT data, while the bottom row shows the Mid-RT data collected 3 weeks after the radiotherapy. It is seen that,
\begin{itemize}
    \item The overall SUV level decreases 3 weeks after the radiotherapy;
    \item The change in SUV (i.e., tumor's response to radiotherapy) varies over space. In particular, the SUV level gradually decreases from the centroid to the surface of a tumor.
\end{itemize}

Hence, if we let $Y(\bm{s})$ in (\ref{eq:model}) represent the change in SUV level at location $\bm{s}$ within the spatial domain $\mathcal{S}$ occupied by the tumor, then, a statistical spatial model is needed that enables clinicians to understand how $Y(\bm{s})$ depends on a vector of features (including geometric features, therapy dosage, and so on) for treatment optimization and control. However, in practice, the complex relationship between the response $Y$ and features $\bm{x}$ can hardly be directly specified. The inevitably nonlinear and interaction effects motivate us to investigate the non-parametric tree-based methods in this paper. More detailed discussions and rationales are provided in the next section.  

\vspace{8pt}
\subsection{Literature Review and Contributions}
The pioneering work of statistical modeling of spatial data can be found in \cite{Banerjee2004}, \cite{Schabenberger2005} and \cite{Cressie2011}. The mainstay approach, also known as the geostatistical paradigm, models spatial processes by random fields with fully specified covariance functions. 
A linear relationship between covariates and process mean is often assumed, and the model parameters can be obtained through Generalized Least Squares or Maximum Likelihood Estimation. 
The covariance structures are typically derived from moments of multivariate probability distributions and motivated by considerations of mathematical convenience (e.g., stationary, isotropic, space-time separable, etc.).
%
Hence, there have been prolonged research interests to provide flexible and effective ways to construct non-stationary covariance functions \citep{Cressie1999, Gneiting2002, Fuentes2005, Gneiting2006, Ghosh2010, Reich2011, Lenzi2019}. 
The geostatistical modeling paradigm, which heavily relies on random fields, becomes less practical for large problems and 
approximations are commonly used, such as the Gaussian Markov Random Fields representation \citep{Lindgren2011}, Nearest-Neighbor Gaussian Process \citep{ Datta2016, Banerjee2017}, kernel convolution \citep{Higdon1998}, low-rank representation \citep{Cressie2002, Nychka2002, Banerjee2008}, the approximation of likelihood functions \citep{Stein2004, Fuentes2007, Guinness2015}, Bayesian inference for latent Gaussian models based on the integrated nested Laplace
approximations \citep{Rue2009, R-Inla2019}, Lagrangian spatio-temporal covariance function \citep{Gneiting2006}, matrix-free state-space model \citep{Mondal2019}, Vecchia approximations of Gaussian processes \citep{Katzfuss2019}, as well as the multi-resolution approximation (\textit{M}-RA) of Gaussian processes observed at irregular spatial locations \citep{Katzfuss2017}. 

Other spatial models have also been proposed in the literature. Notably, the Markov Random Fields (MRF) model focuses on the (conditional) distribution of the quantity at a particular spatial location given the neighboring observations, such as the auto Poisson model \citep{Besag1974}, Conditional Autoregressive Model \citep{Carlin2003, Liu2018}, etc. The Spatial Moving Average (SMA) approach models a spatial process through a process convolution with a convolution kernel \citep{Higdon1998, Brown2000, Liu2016}. There is also a large body of literature focusing on spatio-temporal data. For example, the Hierarchical Dynamical Spatio-Temporal Models (DSTM) \citep{Wikle1999, Berliner2003, Cressie2011, Stroud2010, Katzfuss2019}, and the SPDE-based modeling approach that aims to integrate governing physics into statistical spatio-temporal models \citep{Brown2000, Hooten2008, Stroud2010, Sigrist2015, Liu2020}. A summary of the latest advances in the spatial modeling with SPDE can be found in \cite{Cressie2011} and \cite{Krainski2019}.

For spatial models in the form of (\ref{eq:model}), one challenge arising from practice is to specify the relationship between the covariates $\bm{x}$ and response $Y$, which can rarely be adequately captured by linear models. For the FDG-PET imaging data presented in Section \ref{sec:motivating}, for example, both non-linear and interaction effects are expected between tumor's response and covariates (such as treatment, geometric features of the tumor, etc.). Hence, non-parametric approaches, especially the additive-tree-based approaches, provide some major modeling advantages. Constructing a tree does not require parametric assumptions on the complex relationship between features and event processes. An individual tree performs a partition of the feature space. For each sub feature space, a predicted value is found for the individuals over that sub feature space. A sum-of-trees model consists of multivariate components that effectively handle the complex interaction effects among features \citep{Chipman2010}. Feature selection is also possible under the framework of additive-tree-based models \citep{Hastie2009, Liu2020b}.

Among the additive-tree-based methods, gradient boosted trees have 
become one of the most successful statistical learning approaches over the last two decades, generating 17 winning solutions among 29 Kaggle challenges in 2015 \citep{Chen2016}. \cite{Schapire1999} introduced the first applicable Boosting method. The main idea of gradient boosting hinges on fitting a sequence of correlated ``weaker learners'' (such as simple trees). Each tree explains only a small amount of variation not captured by previous trees \citep{Hastie2009, Chipman2010}. However, many existing boosting trees, such as XGBoost, do not consider the possible spatial correlation when they are applied to spatial data. To our best knowledge, \cite{Sigrist2020} recently proposed the only boosted-trees-based approach for Gaussian Process and mixed effects model (which captures correlated errors). Such a method minimizes the negative log-likelihood at each tree node splitting, and is available in the R package, GPBoost. The package also provides a range of regularization and tuning parameter options. In our paper, on the other hand, regularizations have been directly added to the objective function in order to control the complexity of individual trees (i.e., number of leaves and leaf weights), leading to a regularized optimziation problem at each tree node splitting. The regularization terms are motivated by the fundamental idea behind boosting trees which involves a sequence of correlated simple trees. This idea has been adopted by XGBoost \citep{Chen2016}, while an alternative approach adopted by the Bayesian Additive Regression Trees (BART) involves assigning prior distributions on parameters charactering the tree structure \citep{Chipman2010}. Hence, the \textbf{main contribution} of the paper is to propose a computationally-efficient gradient boosting method for growing the ensemble trees $\{f_k(\bm{x})\}_{k=1}^{K}$ in (\ref{eq:Z}) for spatially-correlated data. Each tree is grown by solving a regularized optimization problem, where the objective function involves regularizations on tree complexity and takes into account the underlying spatial correlation. The proposed algorithm is referred to as Boost-S, which stands for Gradient Boosted Trees for Spatial Data with covariate information (Boost-S). Boost-S integrates the spatial correlation structure into the classical framework of gradient boosted trees.  In Statistics, Ordinary Least Squares is extended to Generalized Least Squares for correlated data. \textbf{An analogous notion can be formulated here when extending the classical framework of gradient boosted trees to spatially-correlated data, giving rise to the proposed Boost-S}. 

The rest of this paper is structured as follows: Section \ref{sec:Boost-S} presents the technical details of the Boost-S algorithm. The applications and numerical illustrations of Boost-S are presented in Section \ref{sec:numerical}. Section \ref{sec:conclusion} concludes the paper. 

\vspace{8pt}
\section{Boost-S: Technical Details} \label{sec:Boost-S}
This section provides the technical details behind Boost-S. 
Suppose that data are collected from a number of $n$ spatial locations, $\bm{s}_1, \bm{s}_2, ...,\bm{s}_n$. At each location, we observe a response $y$ and a $m$-dimensional feature vector $\bm{x}=(x_1, x_2, ... , x_m)^T$. 
From (\ref{eq:model}) and (\ref{eq:Z}), Boost-S aims to construct an ensemble of binary trees, $\{f_k(\bm{x})\}_{k=1}^{K}$, using gradient boosting such that
\begin{equation} \label{eq:model2}
    Y(\bm{s}) = \sum_{k=1}^{K} f_k(\bm{x}) + \varepsilon(\bm{s}), \quad\quad \bm{s}\in \{\bm{s}_1, \bm{s}_2, ...,\bm{s}_n\}.
\end{equation}

Let $\bm{Y} = (Y(\bm{s}_1),Y(\bm{s}_2),...,Y(\bm{s}_n))^T$ be a multivariate random vector representing the responses from the $n$ spatial locations, and let $\bm{f}^{(k)}$ be a vector of predicted values at the $n$ spatial locations generated from the $k$th tree ($k\geq 0$), we re-write (\ref{eq:model2}) as
\begin{equation} \label{eq:model3}
\bm{Y}= \sum_{k=0}^{K} \bm{f}^{(k)} + \bm{\varepsilon}, \quad\quad \bm{\varepsilon}\sim \mathcal{N}(\bm{0}, \bm{\Sigma}_{\bm{\theta}}).
\end{equation}
where $\bm{f}^{(0)}$ is a vector of zeros corresponding to the initial condition when no tree has been grown. 

 \vspace{8pt}
\subsection{A Regularized Problem}
This subsection presents the detailed tree structures and formulates a regularized optimization problem that leads to a sequence of ensemble trees. 
In (\ref{eq:model2}), each tree $f_k$ is a Classification and Regression Tree (CART) that resides in a binary tree space
\begin{equation}
\mathcal{F} = \left\{f(\bm{x}) = w_{q(\bm{x})}\right\}
\end{equation}
where $q: \mathcal{R}^p \rightarrow T$ and $w \in \mathcal{R}^T$. Here, $T$ represents the number of tree leaves (i.e., terminal nodes),  $w$ is the value on a tree leaf (i.e., leaf weight), and $q$ determines the tree structure (i.e., a mapping that links a feature vector $\bm{x}$ to a tree leaf).  

Suppose that a number of $k-1$ trees have been grown ($k \geq 1$). Then, the (ensemble) predicted values at the $n$ spatial locations are given by $\hat{\bm{y}}^{(k-1)}=\sum_{j=0}^{k-1}\bm{f}^{(j)}$ from the $k-1$ trees. An immediate next step is to construct the $k$th tree and add the new tree to the ensemble such that $\hat{\bm{y}}^{(k)}=\hat{\bm{y}}^{(k-1)}+\bm{f}^{(k)}=\sum_{j=0}^{k}\bm{f}^{(j)}$. For binary trees, this involves finding the optimal split features as well as the split points for the $k$th tree. 
This task can be formulated as a regularized optimization problem:
\begin{equation}
    \min_{\bm{f}^{(k)}}  \left\{  \ell(\hat{\bm{y}}^{(k-1)} + \bm{f}^{(k)}) + \Omega(\bm{f}^{(k)}) \right\} \label{eq:optim}
\end{equation}
where $\ell$ is a loss function that depends on the output of the $k$th tree, and the regularization $\Omega$ is given by:
\begin{equation} \label{eq:omega}
\Omega(\bm{f}) = \gamma T + \frac{1}{2}\lambda\|\bm{w}\|^2. 
\end{equation}

The first term, $\gamma T$, regularizes the depth of the tree (by penalizing the total number of leaves), while the second term is used to regularize the contribution of tree $k$ to the ensemble predictions (by penalizing the weights on tree leaves). Recall that, the fundamental idea behind boosting trees is to construct a sequence of correlated ``weaker learners'' (i.e., simple trees), where each ``weaker learner'' is added to explain the unexplained variation by existing trees in the ensemble \citep{Chipman2010}. Hence, the regularization (\ref{eq:omega}) effectively controls the complexity of individual trees. In fact, it is worth noting that the regularization also helps to prevent the well-known overfitting issue of boosting trees. When a sufficient number of trees have been included in the ensemble, the penalty of adding one more tree may dominate the benefit (of adding more trees), which stops the algorithm from growing more trees. 

Because the multivariate response $\bm{Y}$ is spatially correlated with the covariance matrix $\bm{\Sigma}_{\bm{\theta}}$, a sensible choice for the loss function $\ell$ is the squared Mahalanobis length of the residual vector:
\begin{equation}
\begin{split}
      \ell(\hat{\bm{y}}^{(k-1)}+\bm{f}^{(k)}) & = \ell(\hat{\bm{y}}^{(k)}) \\
      &\equiv (\bm{y}-\hat{\bm{y}}^{(k)})^{T}\bm{\Sigma}^{-1}_\theta(\bm{y}-\hat{\bm{y}}^{(k)}) 
\end{split}
    \label{eq:mahalanobis}
\end{equation}
and (\ref{eq:optim}) can thus be written as
\begin{equation} \label{eq:optim2}
\min_{\bm{f}^{(k)}}  \left\{  (\bm{y}-\hat{\bm{y}}^{(k)})^{T}\bm{\Sigma}^{-1}_\theta(\bm{y}-\hat{\bm{y}}^{(k)}) + \gamma T + \frac{1}{2}\lambda\|\bm{w}\|^2 \right\}. 
\end{equation}

Note that, (\ref{eq:optim2}) above extends the classical XGBoost which does not consider the correlation among the elements of $\bm{Y}$ \citep{Chen2016}. The extension made by this paper is in analogy to the extension from Ordinary Least Squares to Generalized Least Squares. However, such an extension requires new algorithms for the problem (\ref{eq:optim2}) to be efficiently solved. 

The regularized optimization above is a formidable combinatorial optimization problem which can hardly be directly solved. Hence, we approximate the objective function $\ell(\hat{\bm{y}}^{(k)}) + \Omega(\bm{f}^{(k)})$ by a
second-order multivariate Taylor expansion: 
\begin{equation}
\begin{split}
    \ell(\hat{\bm{y}}^{(k)}) &+  \Omega(\bm{f}^{(k)})  \\ &\approx  \ell(\hat{\bm{y}}^{(k-1)}) + \bm{g}^T \bm{f}^{(k)} + \frac{1}{2}(\bm{f}^{(k)})^T \bm{H} \bm{f}^{(k)} + \Omega(\bm{f}^{(k)})
\end{split}
\label{taylor}
\end{equation}
where $\bm{g}$ is the column gradient vector of the loss function with its $i$th element given by $g_i = \partial \ell(\hat{\bm{y}}^{(k-1)}) / \partial\hat{y}_i^{(k-1)}$, and $\bm{H}$ is the Hessian matrix with its $(i,j)$th entry being given by $h_{i,j} = \partial^2 \ell(\hat{\bm{y}}^{(k-1)}) / \partial\hat{y}_i^{(k-1)}\partial\hat{y}_j^{(k-1)}$. Because the first term on the right-hand-side of (\ref{taylor}) is a constant, it is sufficient to minimize the sum of the remaining three terms:
\begin{equation}
    L^{(k)} =   \bm{g}^T \bm{f}^{(k)} + \frac{1}{2}(\bm{f}^{(k)})^T \bm{H} \bm{f}^{(k)} + \Omega(\bm{f}^{(k)}).
\label{eq:taylor2}
\end{equation}

Note that, for any given tree structure, it is possible to define a set $I_p = \{i\mid q(x_i) = p\}$ that consists of all samples that fall into leaf $p$. Then, we let $\bm{g}_p$ be a column vector that only retains the elements in $\bm{g}$ corresponding to samples in $I_p$, and similarly, let $\bm{H}_{p,q}$ be a matrix by only keeping the rows and columns of $\bm{H}$ corresponding to samples in $I_p$ and $I_q$, respectively.

\begin{figure}[h!]
	\vspace{-2mm}
	\centerline{\includegraphics[width=0.7\linewidth]{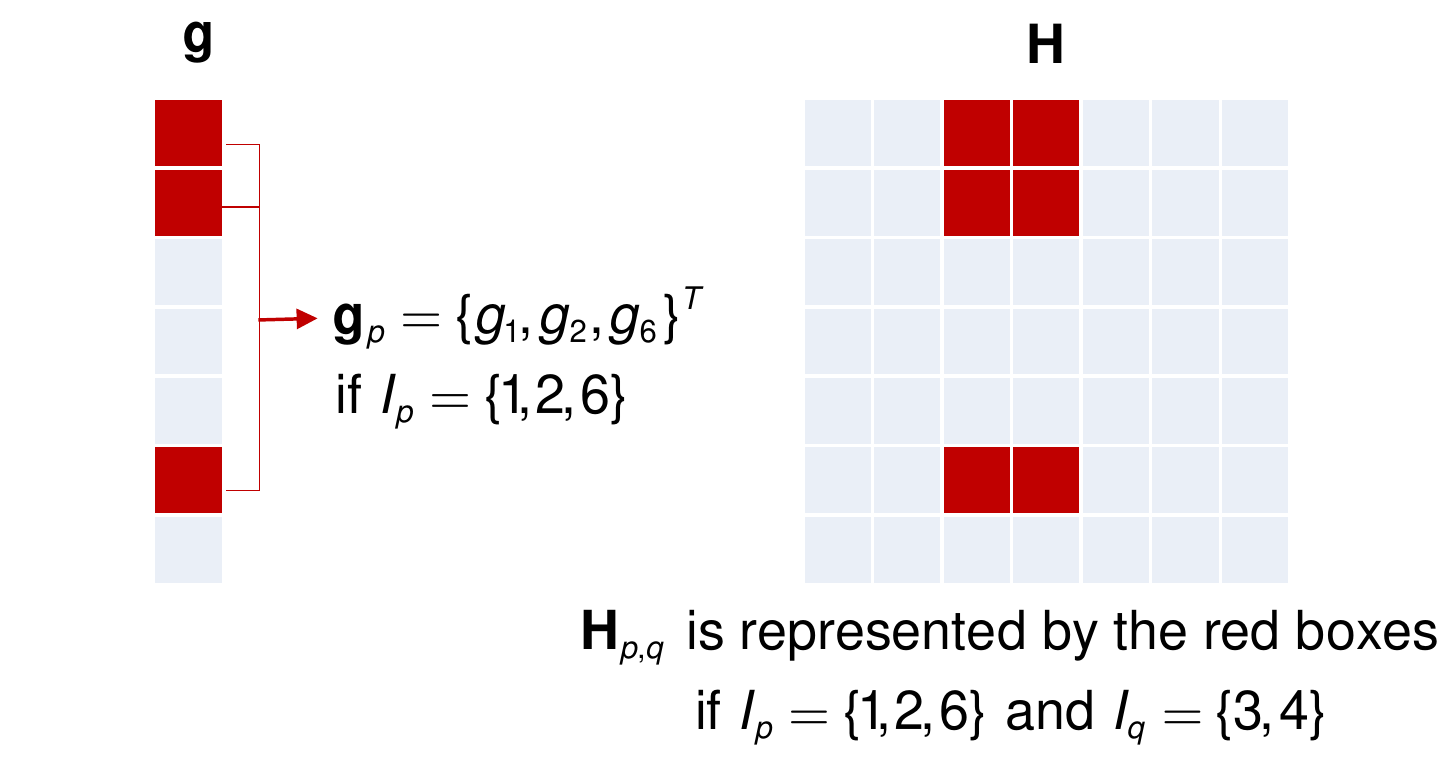}}
	\vspace{-3mm}
	\caption{An illustration of the vector $\bm{g}_p$ and the matrix $\bm{H}_{p,q}$}
	\label{fig:gH}
\end{figure}

Figure \ref{fig:gH} provides an illustration of how $\bm{g}_p$ and $\bm{H}_{p,q}$ are constructed. Consider a simple example where $n=7$ (i.e., only 7 samples are available), then, the dimensions of the gradient vector $\bm{g}$ and the Hessian matrix $\bm{H}$ are $7\times 1$ and $7\times 7$, respectively. Suppose that $I_p={1,2,6}$ and $I_q={3,4}$. Then, the vector $\bm{g}_p$ consists of the 1st, 2nd and the 6th element in  $\bm{g}$, and the matrix $\bm{H}_{p,q}$ is a $3\times 2$ matrix that retains the entries $\bm{H}$ located at the intersections of rows 1, 2 and 6, and  columns 3 and 4. 

Then, for any given tree structure $q(\bm{x})$, we re-write (\ref{eq:taylor2}) as follows:
\begin{equation}
\begin{aligned}
    L^{(k)} = \sum_{p = 1}^{T} \bm{g}^{T}_p \bm{w}_p + \frac{1}{2}\left\{\sum_{p\in T} \bm{w}^{T}_p \bm{H}_{p,p} \bm{w}_p\right\}\\ + \frac{1}{2}\left\{\sum_{(p,q)\in C_2^{T}} \bm{w}^{T}_p \bm{H}_{p,q} \bm{w}_q\right\} + \Omega(\bm{f}^{(k)})
\end{aligned}
\label{eq:14}
\end{equation}
where $C_2^{T}$ denotes the combination (i.e., the number of 2-combinations from a given set of $T$ elements), $\bm{w}_p = w_p \bm{1}_{|I_p|}$ with $w_p$ and $\bm{1}_{|I_p|}$ respectively being the weight on tree leaf $p$ and a column vector of ones of dimension $|I_p|$, and $|\cdot|$ represents the cardinality of a set. 

Substituting (\ref{eq:omega}) into \eqref{eq:14} yields
\begin{equation} \label{eq:Lfinal}
\begin{split}
            L^{(k)} = &\gamma T + \sum_{p = 1}^{T}\left\{ w_p\sum_{i \in I_p}g_i + \frac{1}{2}\left[\lambda+\sum_{(i,j) \in I_p}h_{i,j}\right]w_p^2 \right\} \\ 
            & + \frac{1}{2}\sum_{p = 1}^{T}\left\{\sum_{q=1; q\neq p}^{T}\left[\lambda+\sum_{i \in I_p; j \in I_q}h_{i,j} \right] w_p w_q  \right\}. 
\end{split}
\end{equation}


Taking the partial derivatives of (\ref{eq:Lfinal}) with respect to $\{w_p\}_{p=1}^T$ leads to a system of equations, which provides the key \textit{computational advantag}e: given any tree structure, the optimal weights $\bm{w}=\{w_1, w_2, ...,w_T\}$ can be quickly found by solving the linear system:
\begin{equation} \label{eq:linear}
\Xi \bm{w} = - \tilde{\bm{g}}
\end{equation}
where $\Xi$ is a $T \times T$ matrix with its $p$th row given by 
\begin{equation}
    \frac{1}{2}(\lambda + \sum_{i\in I_p;j\in I_1}h_{i,1} ), \frac{1}{2} (\lambda + \sum_{i\in I_p;j\in I_2}h_{i,2} ),...,
    (\lambda + \sum_{(i,j)\in I_p}h_{i,j}),...,
    \frac{1}{2} (\lambda + \sum_{i\in I_p;j\in I_T}h_{i,T} )
\end{equation}
and $\tilde{\bm{g}}$ is a $T \times 1$ vector with its $p$th element given by $\sum_{i \in I_p}g_i $. 

Obtaining the linear system (\ref{eq:linear}) plays an extremely important role in searching for the optimal tree: given any candidate tree structure, it is possible to quickly and accurately find the optimal weights $\bm{w}$ on the leaf nodes by solving (\ref{eq:linear}) using least squares, i.e., no numerical search is required. By substituting the optimal $\bm{w}$ into (\ref{eq:taylor2}) immediately yields the value of the objective function $L^{(k)}$. 

\vspace{8pt}
\subsection{The Sequential Update of the Unknown Covairance Matrix}
Constructing the linear system (\ref{eq:linear}) and evaluating the objective function $L^{(k)}$ require a known covariance matrix of the errors,  $\bm{\Sigma}_{\bm{\theta}}$. However, $\bm{\Sigma}_{\bm{\theta}}$ is not known and needs to be estimated before the first tree ($k=1$), or any subsequent tree ($k>1$), can be constructed. In this section, we describe how $\bm{\Sigma}_{\bm{\theta}}$ can be consistently estimated before the $k$th tree is constructed, given the outputs from the first $k-1$ trees. 

Suppose that $k-1$ trees have been constructed ($k\geq 1$), and let $\bm{r}^{(k-1)} = \bm{Y}-\sum \bm{f}^{(k-1)}$ be the residual vector. It follows from (\ref{eq:model3}) that $\bm{r}^{(k-1)}$ is Gaussian with the covariance $\bm{\Sigma}_{\bm{\theta}}$. Note that, the mean of $\bm{r}^{(k-1)}$ may not even be close to zero when $k$ is small, i.e., when there are not sufficient trees in the ensemble to well capture the mean of $\bm{Y}$. In this case, we model $\bm{r}^{(k-1)}$ by  a Locally Weighted Mixture of Linear Regressions (LWMLR) \citep{Stroud2001}:
\begin{equation} \label{eq:LWMLR}
\bm{r}^{(k-1)} = \sum_{j=1}^{J} \pi^{T}_j(\bm{s}) \bm{k}_j(\bm{s})\bm{\beta}_j + \varepsilon(\bm{s}), \quad\quad k\geq1
\end{equation}
where $\bm{k}_j(\bm{s})=\{k_{j1}(\bm{s}),...,k_{jq}(\bm{s})\}$ is a set of spatial basis functions, 
$\bm{\beta}_j=(\beta_{j1},...,\beta_{jq})$ is a vector of unknown coefficients, $\pi_j(\bm{s})$ is a Gaussian kernel given as follows:
\begin{equation}
    \pi_j(\bm{s}) \propto  |\bm{V}_j|^{-1/2}\exp\left\{-\frac{1}{2}(\bm{s}-\bm{\mu}_j)^T\bm{V}_j^{-1}(\bm{s}-\bm{\mu}_j) \right\}
\end{equation}

Note that, we may re-write (\ref{eq:LWMLR}) as a linear model,  $\bm{r}^{(k-1)}=\bm{X}\bm{B}+\bm{\varepsilon}$, 
where $\bm{X}=(\text{diag}(\pi_1)\bm{X}_1, ..., \text{diag}(\pi_J)\bm{X}_J)$, $\bm{X}_j=(\bm{k}_j^T(\bm{s}_1), \bm{k}_j^T(\bm{s}_2),...,\bm{k}_J^T(\bm{s}_J))^T$, and $\bm{B} = (\bm{\beta}_1, \bm{\beta}_2, ..., \bm{\beta}_J)^T$.
Then, it is possible to obtain a consistent estimate of $\bm{\Sigma}_{\bm{\theta}}$, $\hat{\bm{\Sigma}}_{\bm{\theta}}^{(k-1)}$, using the Feasible Generalized Least Square (FGLS), before the $k$th tree can be constructed. 
\vspace{8pt}
\subsection{The Boost-S Algorithm}
\begin{figure}[h!]
	\centerline{\includegraphics[width=1\linewidth]{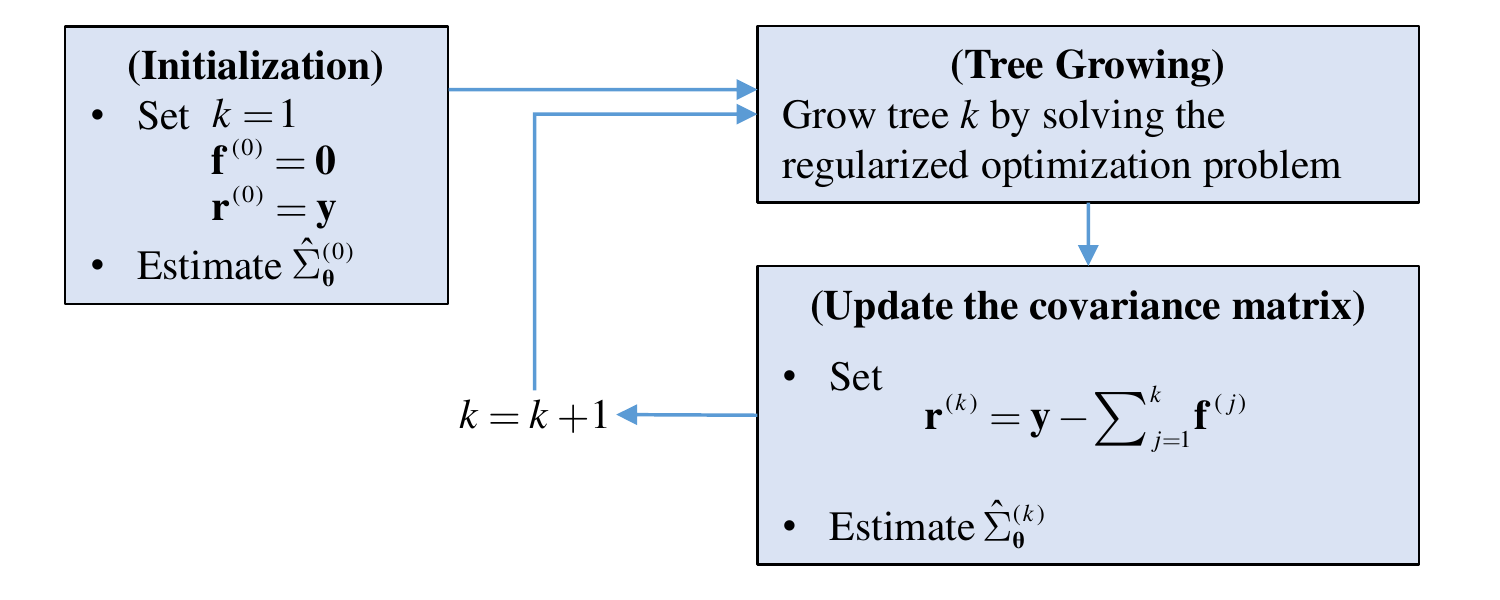}}
	\vspace{-14pt}
	\caption{An illustration of the flow of the Boost-S algorithm}
	\label{fig:graph}
\end{figure}

Following the discussions above, Figure \ref{fig:graph} provides a high-level illustration of the flow of the Boost-S algorithm. At the initialization stage, we obtain the initial estimate of the covariance matrix $\hat{\bm{\Sigma}}_{\bm{\theta}}^{(0)}$. Then, by solving the regularized optimization problem (\ref{eq:optim}), a new tree $k$ is constructed and added to the ensemble. Next, the estimate of the covariance matrix is updated $\hat{\bm{\Sigma}}_{\bm{\theta}}^{(k)}$ before tree $k+1$ can be constructed. The steps are repeated until $K$ trees have been grown in the ensemble. 
The Boost-S algorithm is formalized by Algorithm 1. 

\begin{algorithm}
	\DontPrintSemicolon
	\SetAlgoLined
	\SetKwInOut{Data}{Data}\SetKwInOut{Output}{Output}\SetKwInOut{Preprocesses}{Preprocesses}
	\SetKwInOut{Begin}{Begin}\SetKwInOut{Parameters}{Parameters}\SetKwInOut{Input}{Input}
	
	Set the values for $\lambda$, $\gamma$ and $K$
	
	Let k = 0, $\bm{f}^{(0)}=\bm{0}$ and $\bm{r}^{(0)}=\bm{y}$
	
	Obtain the initial estimate,  $\hat{\bm{\Sigma}}_{\bm{\theta}}^{(0)}$, from (\ref{eq:LWMLR}) using FGLS
	
	\For{k=1,...,K} 
	{ Grow tree $k$ by repeating the following steps:\\
		(\textbf{i}) Given the current tree topology, generate a set of all possible new tree structures by splitting a tree node based on candidate split variables and  candidate split values. \\
		(\textbf{ii}) For each new tree structure, obtain the weights $\bm{\omega}$ on the leaf nodes by solving the linear system $\Xi \bm{w} = - \tilde{\bm{g}}$, 
		and evaluate the objective function $L^{(k)}$ in (\ref{eq:taylor2}) for the new topology. \\
		(\textbf{iii}) If there exists at least one new tree structure that further reduces the objective function (over the existing tree structure), retain the new topology that generates the greatest reduction and go to (\textbf{i}); otherwise, terminate the tree growing process for tree $k$, and go to the next step.\\
		(\textbf{iv}) Update the estimate $\hat{\bm{\Sigma}}_{\bm{\theta}}^{(k)}$using FGLS.}
	\label{algorithm}
	\caption{Boost-S: Gradient Boosted Trees for Spatial Data}\label{Boost-S}
\end{algorithm}

\vspace{-12pt}


\section{Illustration of Boost-S on the FDG-PET Imaging Data} \label{sec:numerical}
This section re-visits the motivating application presented in Section \ref{sec:motivating}. We apply Boost-S to real datasets and compare the performance of Boost-S to that of existing approaches.  

\vspace{8pt}
\subsection{Data}
The data used in this section are obtained from 25 patients diagnosed with locally advanced and surgically unresctable non-small cell lung cancer (NSCLC) who enrolled onto the FLARE-RT clinical trial (NCT02773238). For each patient, this clinical trial data set contains geographic features, dosage, Pre-RT and Mid-RT SUV levels. As discussed in Section \ref{sec:motivating}, the goal is to model the difference between the Pre-RT SUV (before treatment) and Mid-RT SUV (during third week of treatment course), which helps clinicians further optimize or control the treatment plans. 

Let $Y(\bm{s})$ represent the ratio between Mid-RT SUV and Pre-RT SUV at location $\bm{s}$ (a voxel in the image) within the spatial domain $\mathcal{S}$ occupied by the tumor, i.e., 
\begin{equation}
    Y(\bm{s}) = \frac{\text{Mid-RT SUV}}{\text{Pre-RT SUV}}. 
    \label{eq:Y}
\end{equation}
If the treatment is effective,  the Mid-RT SUV is expected to be lower than the Pre-RT SUV level. Hence, a lower ratio indicates more effective treatment. 

\vspace{8pt}
\subsection{Application of Boost-S}
We first demonstrate the application of Boost-S on the data collected from one of the 25 patients. The PET scan for this patient has 3110 voxels (i.e., the number of spatial locations). We randomly split the data into two parts: 15\% for training while 85\% for testing. Such a low training-testing ratio is chosen to demonstrate the out-of-sample prediction capability of Boost-S constructed from a relatively small training dataset.

\begin{figure} 
	\centering
	\includegraphics[width=0.6\linewidth]{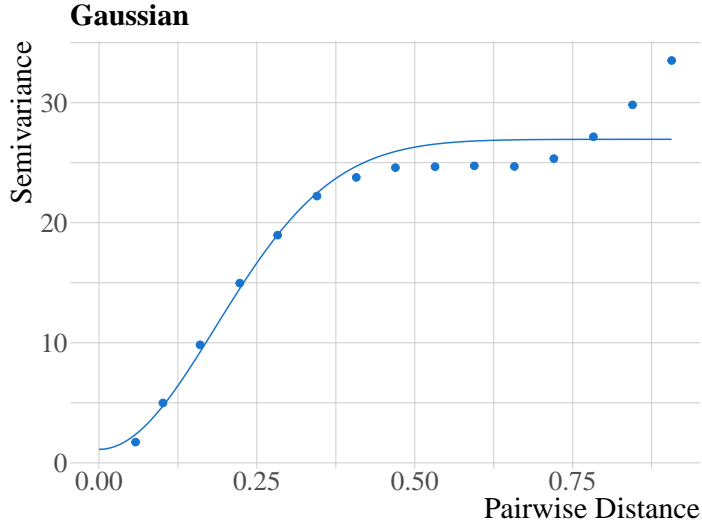}
	\vspace{-12pt}
	\caption{Exploratory analysis on the covariance structure: plot of the empirical and fitted semivariance using FGLS}
	\label{fig:covariance}
\end{figure}

To obtain the initial estimate of the covariance matrix $\hat{\bm{\Sigma}}_{\bm{\theta}}^{(0)}$, the FGLS is used to solve the linear model (\ref{eq:LWMLR}) in Algorithm 1. 
Before the FGLS is performed, one needs to first choose a parametric spatial covariance function $c(\cdot)$ of the process $\varepsilon$ in (\ref{eq:LWMLR}). 
Figure \ref{fig:covariance} shows both the empirical semivariance and the fitted semivariance using FGLS assuming the Gaussian covariance function.  
The sill, range and nugget effect can be clearly seen from Figure \ref{fig:covariance}, and the Gaussian covariance function appears to be an appropriate choice. 

\begin{figure}[h]
	\centerline{\includegraphics[width=0.75\linewidth]{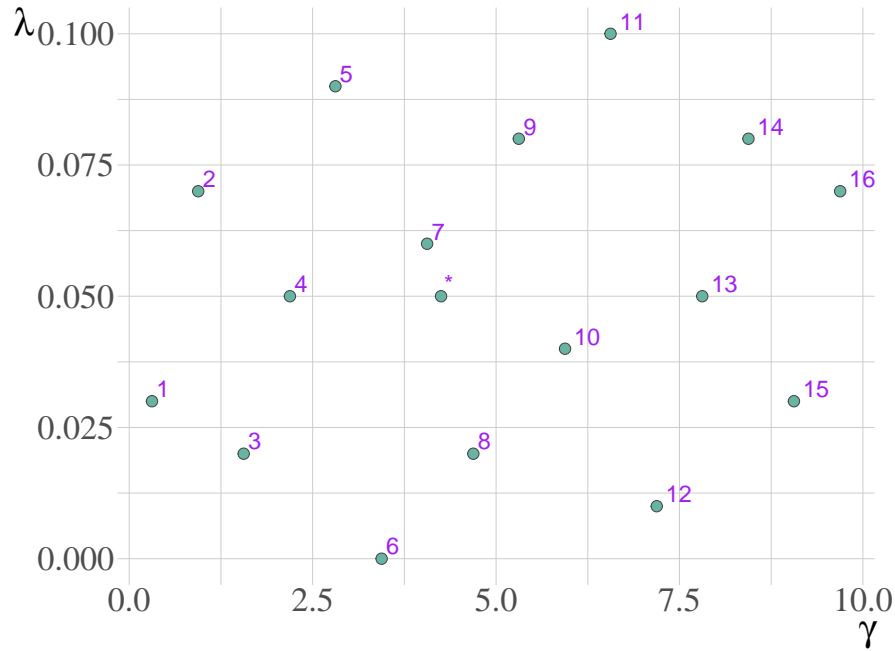}}
	\vspace{-16pt}
	\caption{Maximum projection design of 16 runs with different combinations of $\lambda$ and $\gamma$. Design ``$*$''  yields an appropriate combination such that $\lambda=0.05$ and $\gamma=4.25$.}
	\label{fig:MaxProLHD}
\end{figure}

\begin{figure}[h]
	\centerline{\includegraphics[width=0.75\linewidth]{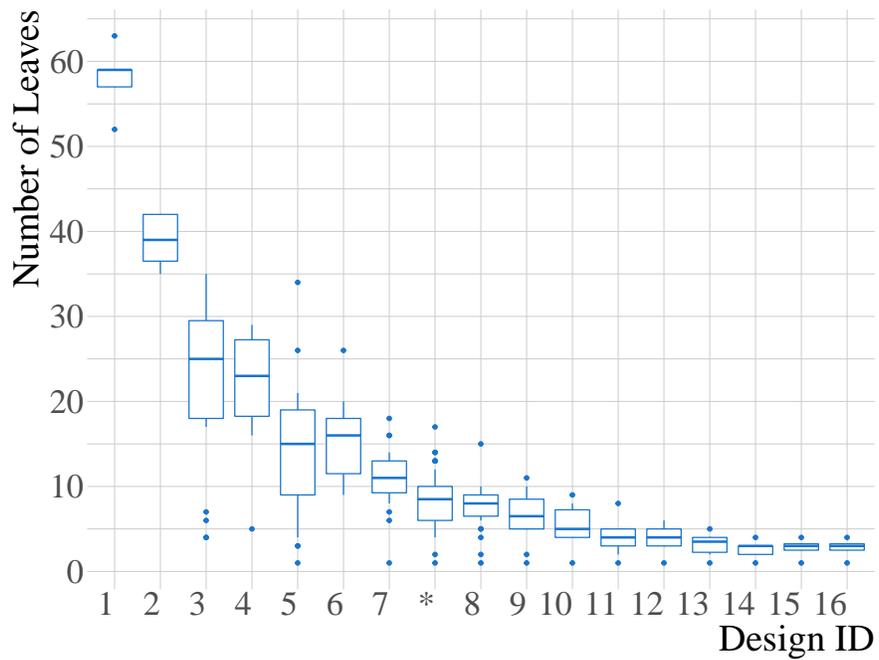}}
	\vspace{-16pt}
	\caption{Boxplot of the number of tree leaves per tree in an ensemble for the 16 candidate designs. Design ``$*$''  indicates the chosen combination such that $\lambda=0.05$ and $\gamma=4.25$.}
	\label{fig:leaves}
\end{figure}

Algorithm 1 requires the tuning parameters $\lambda$ and $\gamma$ to be pre-specified. The choice of these two parameters affects both the depth and contribution of individual trees to the ensemble prediction, which in turn influences the total number of trees in the ensemble. A common strategy suggests that we explore the suitable values for $\lambda$ and $\gamma$ while leaving $K$ as the primary parameter \citep{Hastie2009}. Although it is theoretically possible to perform a grid search for the best combinations of $\lambda$ and $\gamma$ on a two-dimensional space, such an approach may not be practical nor necessary in practice when it is computationally intensive to run Boost-S on big datasets. Hence, we resort to a powerful tool in computer experiments---the space-filling designs \citep{Joseph2016}. The idea of space-filling designs is to have points everywhere in the experimental region with as few gaps as possible, which serves our purpose very well. 
Figure \ref{fig:MaxProLHD} shows the Maximum Projection Latin Hypertube Design (MaxProLHD, \cite{Joseph2015}) of 16 runs with different combinations of $\lambda$ and $\gamma$, where the experimental ranges for these two parameters are respectively $[0,0.1]$ and $[0,10]$. 
For each design, Figure \ref{fig:leaves} shows the box plot of the number of tree leaves per tree in an ensemble for each combination of $\lambda$ and $\gamma$. Since the key idea behind boosting trees is that each individual tree needs to be kept simple \citep{Hastie2009}, we identify that Designs \#7, \#8 and \#9 provide the most suitable combinations of $\lambda$ and $\gamma$. From Figure \ref{fig:MaxProLHD}, these three design points are adjacent to each other, indicating that the appropriate choices for $\lambda$ and $\gamma$ are approximately within $[0.025,0.075]$ and $[4,5.5]$. A refined search in a much smaller experimental region yields an appropriate combination of $\lambda=0.05$ and $\gamma=4.25$ (Design ``$*$'' in Figure \ref{fig:MaxProLHD}). Design ``$*$'' is between Designs \#7 and \#8, and the 25th, 50th and 75th empirical quartiles of the number of tree leaves are  6, 8 and 10 as shown in Figure \ref{fig:leaves}. 
\vspace{-12pt}

\vspace{-46pt}
\clearpage

\begin{figure}[h]
	\centerline{\includegraphics[width=0.9\linewidth]{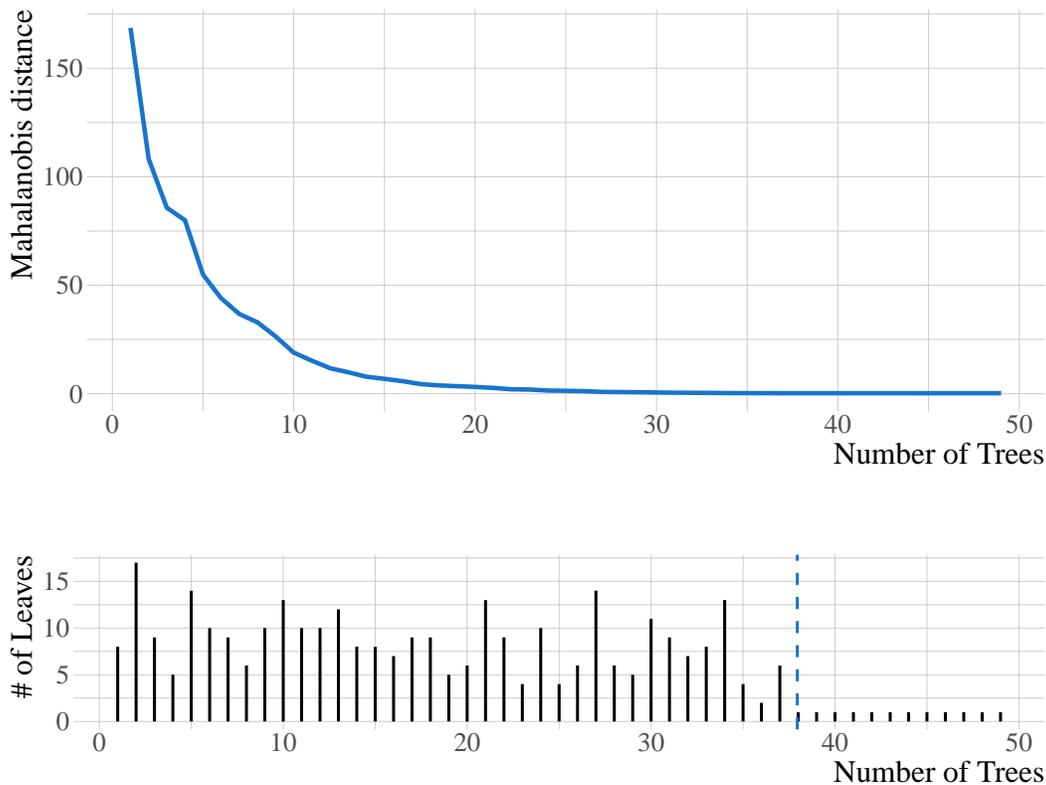}}
	\vspace{-20pt}
	\caption{Top panel: the Mahalanobis distance (i.e., objective function) decreases as the number of trees grows; Bottom panel: the number of tree leaves of individual trees.}
	\label{objective}
\end{figure}

After $\lambda$ and $\gamma$ have been appropriately chosen, 49 trees are constructed to form the ensemble predictor using Algorithm 1.  
Figure \ref{objective} (top panel) shows that the Mahalanobis distance (i.e., the objective function) decreases as more trees have been included into the ensemble, indicating that the algorithm is working as expected. Figure \ref{objective}  (bottom panel) shows the number of tree leaves for individual trees in this ensemble. It is interesting to note that the algorithm no longer splits the (root) tree node after 37 trees have been grown. In addition, the outputs of these one-node trees are all zeros, indicating that all trees after tree 37 are completely redundant. 
This is precisely due to the regularization $\gamma T$ and $\frac{1}{2}\lambda\|\bm{w}\|^2$ in (\ref{eq:optim}). The gain in $\ell$ is outweighted by the loss caused by $\Omega$ if one more tree is added. 

\vspace{-16pt}
\vspace{-12pt}

\begin{figure}[h]
	\centerline{\includegraphics[width=0.8\linewidth]{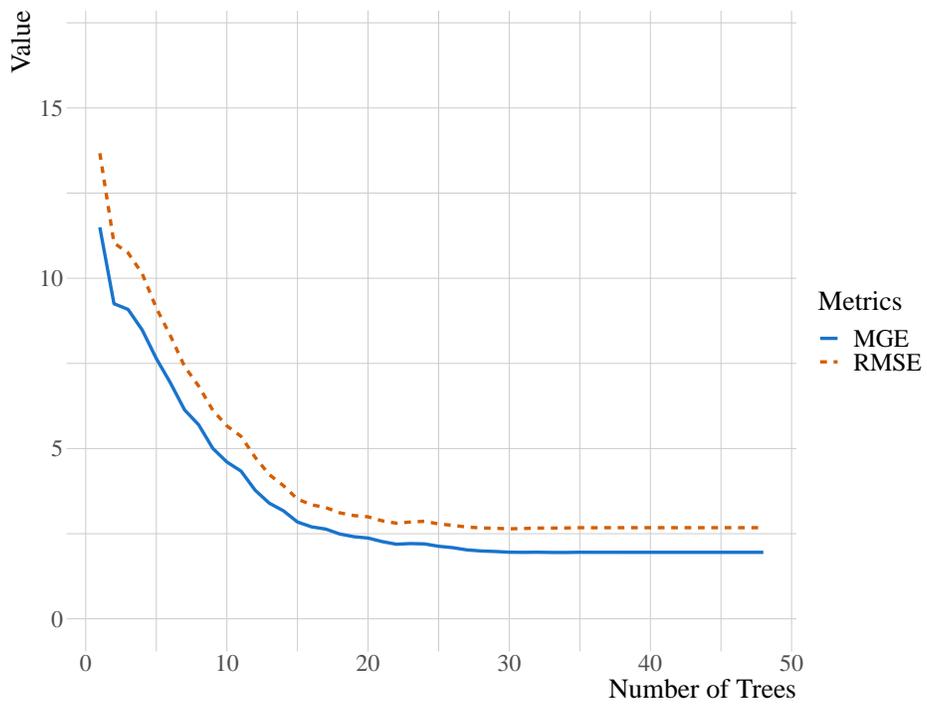}}
	\vspace{-16pt}
	\caption{Out-of-sample model performance in terms of RMSE and MGE}
	\label{assessments}
\end{figure}

\begin{figure}[h]
	\centerline{\includegraphics[width=0.8\linewidth]{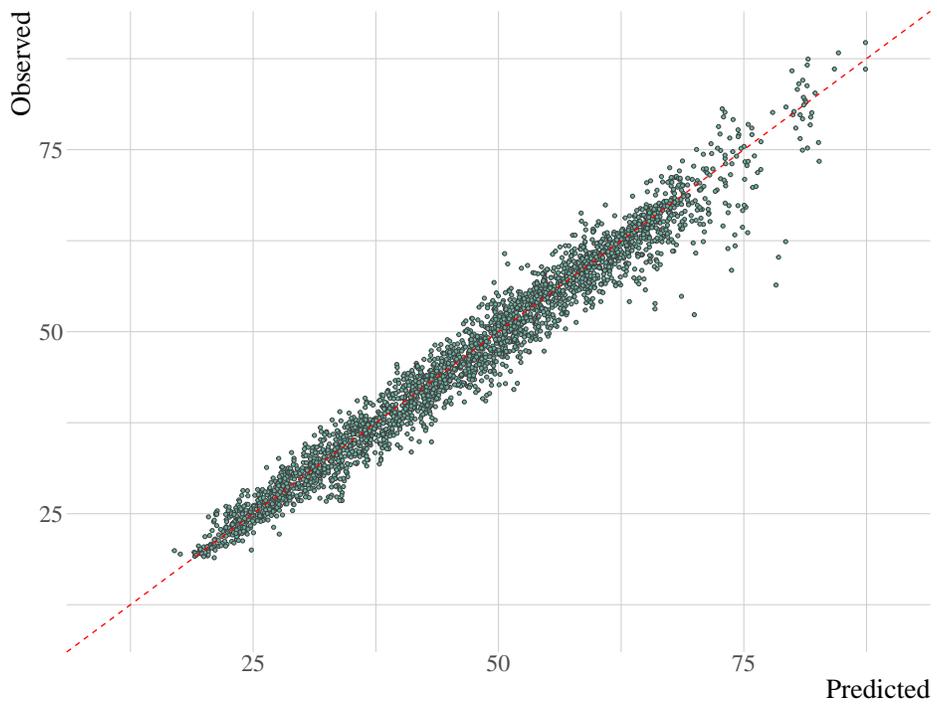}}
	\vspace{-16pt}
	\caption{Out-of-sample predictions against actual observations}
	\label{obs.vs.pred}
\end{figure}

Applying the constructed ensemble trees to the testing dataset, Figure \ref{assessments} shows the out-of-sample Root-Mean-Square-Error (RMSE) and Mean-Gross-Error (MGE). We see that both performance metrics decrease as more trees are included and stabilize approximately after 30 to 40 trees have been grown. As discussed above, this is precisely due to the fact that all trees after \#37 are redundant with zero output. Figure \ref{obs.vs.pred} shows the (out-of-sample) predictions against actual observations of the SUV level at differnt voxels. The figure shows that the proposed Boost-S accurately predicts the SUV levels given the covariate information.   

\vspace{-16pt}

\vspace{-46pt}
\clearpage

\begin{figure}[h]
	\centerline{\includegraphics[width=0.9\linewidth]{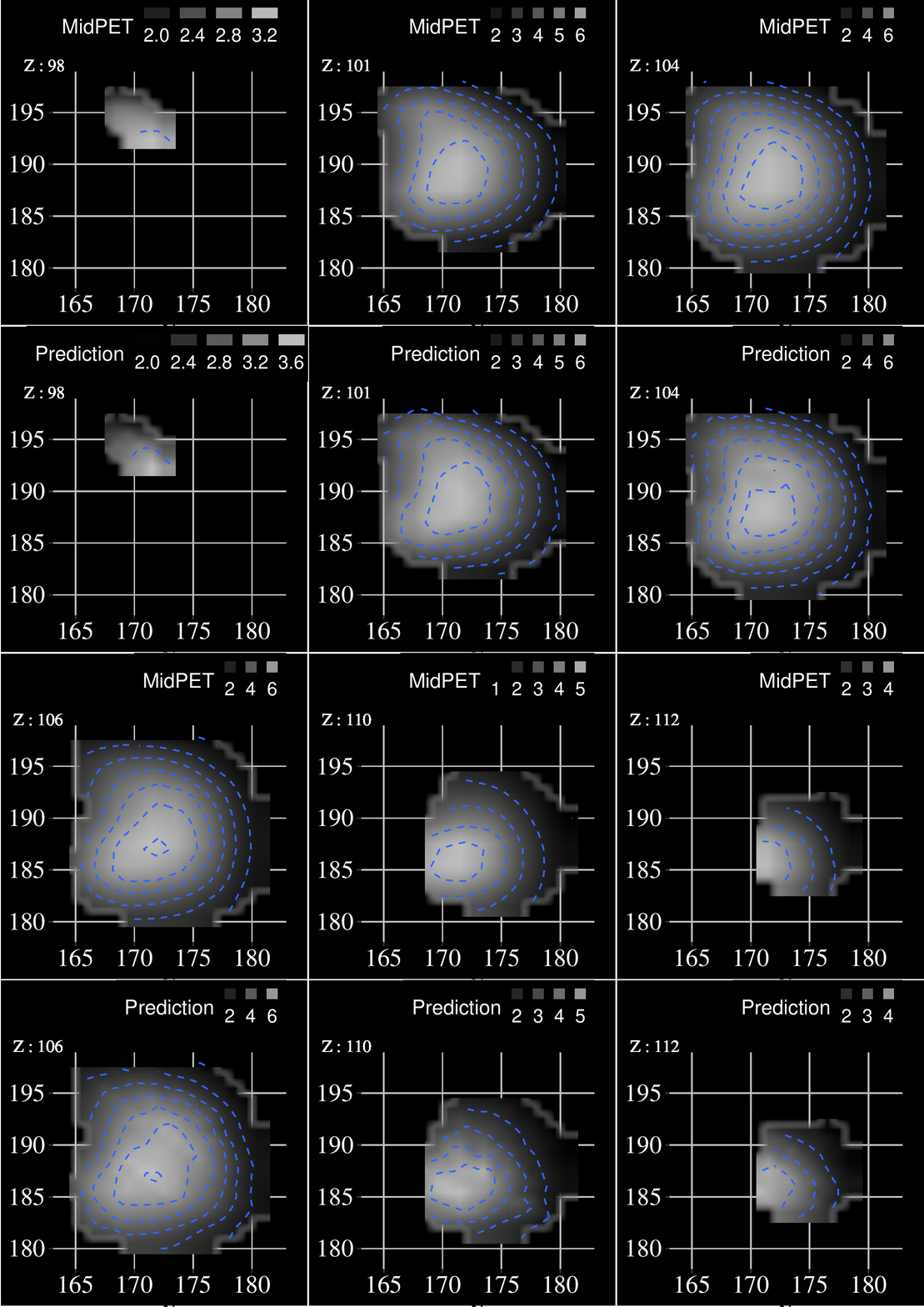}}
	\caption{MidPET images on the $x-y$ plane for different values of $z$. Rows 1 and 3: observed images; Row 2 and 4: resconstructed images.}
	\label{slice}
\end{figure}
\clearpage

Furthermore, we reconstruct the Mid-RT SUV (3 weeks after treatment) by the proposed Boost-S, and compare the actual and predicted Mid-RT SUV using slice plots (Figure \ref{slice}). Because a tumor is a 3D object with three coordinates, $x$, $y$, and $z$, the MidPET images are shown on the $x-y$ plane for given values of $z$. We see that the ensemble trees are capable of predicting the Mid-RT SUV for the entire tumor body. 

\subsection{Comparison Study}
We compare the predictive capability of Boost-S to five other methods using the data collected from 25 patients. The five other methods included in the comparison study are:
Random Forests (RF) \citep{breiman2001random}, Extreme Gradient Boosting Trees (XGBoost) \citep{Chen2016}, non-parametric cubic splines, universal kriging with a linear spatial trend, and multiple linear regression without considering the spatial correlation. For each patient, 15\% of the observations are randomly chosen as the training dataset, and all 6 models are constructed to generate the out-of-sample predictions using the testing dataset. Such a low training-testing ratio is used to test the predictive capabilities of these methods. 

Tables 1, 2 and 3 respectively show the Mean Gross Error (MGE), Relative Error (RE) and Rooted Mean Squared Error (RMSE) of the out-of-sample predictions for 25 patients using 6 different methods. It is interesting to see that
\begin{itemize}
	\vspace{-8pt}
	\item The proposed Boost-S yields the lowest MGE for 19 out of the 25 patients, while RF performs the best for the remaining 6 patients;
		\vspace{-8pt}
	\item The proposed Boost-S yields the lowest RE for 18 out of the 25 patients, while RF performs the best for the remaining 7 patients;
		\vspace{-8pt}
	\item The proposed Boost-S yields the lowest RMSE for 17 out of the 25 patients, while RF performs the best for the remaining 8 patients;
\end{itemize}

\vspace{-8pt}
Hence, we conclude that, \textbf{the proposed Boost-S provides the best performance for majority of the patients in terms of all three performance measures, although no method uniformly outperforms others (which is realistic and expected given the well-known modeling power of RF and XGBoost)}. 
The reason why Boost-S outperforms other additive-tree-based methods, i.e., RF and XGBoost, is because of its capability of accounting for the spatial correlation among observations. The reason why Boost-S outperforms universal Kriging and multiple linear regression is due to the advantages of non-parametric tree-based method in capturing complex non-linear and interaction effects between features and responses. The observation demonstrates the effectiveness of Boost-S as a useful extension to existing ensemble tree-based methods by accounting for the spatial correlation among observations.

\begin{table} \label{table:1}
	\centering
	\caption{Mean Gross Error of the out-of-sample predictions for 25 patients using 6 different methods (note that: rows 1 to 25 respectively show the results corresponding to the 25 patients, while the last row shows the \textit{p}-value of the one-side paired Wilcoxon test)}
	\begin{tabular}{|l|l|l|l|l|l|}
		\hline
		Boost-S & RF & XGBoost & Cubic Splines & Universal Kriging & Linear Regression \\ \hline
		8.24 & \textbf{6.78 }& 7.31 & 8.45 & 10.32 & 9.01 \\ \hline
		10.68 &\textbf{9.61} & 10.55 & 11.57 & 13.41 & 13.79 \\ \hline
		\textbf{5.84} & 6.22 & 6.32 & 7.14 & 10.37 & 9.08 \\ \hline
		\textbf{4.30 }& 4.37 & 4.68 & 6.19 & 6.76 & 6.20 \\ \hline
		\textbf{3.15} & 4.45 & 3.97 & 4.70 & 10.04 & 8.58 \\ \hline
		\textbf{1.95} & 2.40 & 2.51 & 3.07 & 5.42 & 4.25 \\ \hline
	\textbf{5.20} & 5.40 & 5.67 & 5.47 & 13.30 & 10.34 \\ \hline
		\textbf{5.04} & 5.30 & 5.93 & 9.00 & 10.39 & 9.88 \\ \hline
	\textbf{8.59 }& 9.28 & 9.50 & 9.22 & 12.55 & 11.79 \\ \hline
		\textbf{2.37} & 2.79 & 3.17 & 4.46 & 6.70 & 5.97 \\ \hline
		9.08 & \textbf{8.63} & 8.91 & 11.16 & 10.79 & 11.22 \\ \hline
		\textbf{1.44} & 1.74 & 2.08 & 2.64 & 4.00 & 2.99 \\ \hline
		\textbf{1.65} & 1.91 & 2.08 & 3.61 & 4.36 & 3.69 \\ \hline
		\textbf{0.74} & 1.32 & 1.35 & 1.58 & 2.48 & 1.83 \\ \hline
		10.19 & \textbf{9.77} & 10.59 & 10.98 & 15.03 & 13.71 \\ \hline
	\textbf{4.17} & 4.37 & 4.25 & 7.79 & 9.64 & 7.89 \\ \hline
		\textbf{2.71} & 3.20 & 3.78 & 4.22 & 7.15 & 6.47 \\ \hline
		\textbf{6.53} & 7.06 & 8.27 & 8.41 & 13.43 & 12.18 \\ \hline
		13.10 & \textbf{10.62 }& 11.36 & 12.57 & 13.77 & 14.23 \\ \hline
		\textbf{3.51} & 3.96 & 4.06 & 4.44 & 9.30 & 6.85 \\ \hline
	\textbf{1.22} & 1.24 & 1.57 & 1.58 & 2.55 & 2.03 \\ \hline
		\textbf{4.17} & 4.82 & 5.01 & 7.69 & 11.25 & 8.71 \\ \hline
		\textbf{5.47} & 6.62 & 6.43 & 10.65 & 18.82 & 12.05 \\ \hline
		3.87 & \textbf{3.63} & 4.13 & 5.08 & 5.52 & 5.47 \\ \hline
		\textbf{1.86} & 2.42 & 2.58 & 3.57 & 5.60 & 4.19 \\ \hline\hline
		N.A. & 0.04 & 0.001 & $<10^{-6}$ & $<10^{-7}$ & $<10^{-7}$ \\ \hline
	\end{tabular}
\end{table}

\begin{table} \label{table:2}
	\centering
	\caption{Relative Error (in \%) of the out-of-sample predictions for 25 patients using 6 different methods (note that: rows 1 to 25 respectively show the results corresponding to the 25 patients, while the last row shows the \textit{p}-value of the one-side paired Wilcoxon test)}
	\begin{tabular}{|c|c|c|c|c|c|}
		\hline
		Boost-S & RF & XGBoost & Cubic Splines & Universal Kriging & Linear Regression \\ \hline
		9.67 & \textbf{8.01} & 8.58 & 9.91 & 12.29 & 10.68 \\ \hline
		56.62 & \textbf{51.01} & 55.37 & 60.03 & 60.47 & 67.27 \\ \hline
		\textbf{11.04} & 12.52 & 11.93 & 13.72 & 18.40 & 17.01 \\ \hline
		\textbf{5.03} & 5.34 & 5.53 & 7.60 & 8.36 & 7.61 \\ \hline
		\textbf{7.30} & 9.41 & 8.26 & 10.77 & 20.25 & 19.07 \\ \hline
	\textbf{4.42} & 5.73 & 5.65 & 7.29 & 12.52 & 9.72 \\ \hline
		\textbf{8.92} & 9.41 & 9.52 & 9.72 & 20.53 & 18.06 \\ \hline
		23.25 & \textbf{22.32} & 25.52 & 43.34 & 43.51 & 43.78 \\ \hline
		\textbf{13.62} & 15.45 & 15.18 & 14.45 & 20.04 & 18.91 \\ \hline
		\textbf{2.28} & 2.74 & 3.07 & 4.26 & 6.48 & 5.77 \\ \hline
		11.44 & \textbf{10.78} & 10.94 & 14.42 & 13.41 & 14.39 \\ \hline
	\textbf{1.09} & 1.32 & 1.55 & 1.99 & 3.06 & 2.25 \\ \hline
		\textbf{3.66} & 4.25 & 4.55 & 8.00 & 9.70 & 8.20 \\ \hline
		\textbf{1.49} & 2.75 & 2.72 & 3.17 & 4.70 & 3.59 \\ \hline
		21.40 & \textbf{19.97} & 21.21 & 23.26 & 32.02 & 29.69 \\ \hline
		\textbf{8.36} & 8.67 & 8.13 & 15.27 & 20.18 & 15.49 \\ \hline
		\textbf{3.90} & 4.68 & 5.34 & 6.05 & 9.93 & 9.51 \\ \hline
		\textbf{7.46} & 8.79 & 9.97 & 9.99 & 17.54 & 15.31 \\ \hline
		21.70 & \textbf{17.32} & 18.19 & 20.08 & 21.71 & 23.53 \\ \hline
		\textbf{4.02} & 4.58 & 4.60 & 5.05 & 10.15 & 7.95 \\ \hline
		\textbf{1.91} & 1.94 & 2.43 & 2.53 & 4.08 & 3.24 \\ \hline
		\textbf{6.34} & 7.18 & 7.61 & 11.55 & 17.54 & 13.37 \\ \hline
		\textbf{8.10} & 9.55 & 8.73 & 16.69 & 25.45 & 18.96 \\ \hline
		5.24 & \textbf{5.01} & 5.63 & 7.10 & 7.85 & 7.61 \\ \hline
		\textbf{5.07} & 6.59 & 6.86 & 9.72 & 14.92 & 11.26 \\ \hline\hline
		N.A. & 0.09 & 0.004 & $<10^{-6}$ & $<10^{-7}$ & $<10^{-7}$ \\ \hline
	\end{tabular}
\end{table}

\begin{table} \label{table:3}
	\centering
	\caption{Rooted Mean Squared Error of the out-of-sample predictions for 25 patients using 6 different methods (note that: rows 1 to 25 respectively show the results corresponding to the 25 patients, while the last row shows the \textit{p}-value of the one-side paired Wilcoxon test)}
	\begin{tabular}{|c|c|c|c|c|c|}
		\hline
		Boost-S & RF & XGBoost & Cubic Splines & Universal Kriging & Linear Regression \\ \hline
		11.05 & \textbf{9.01} & 9.66 & 10.89 & 13.00 & 11.66 \\ \hline
		16.46 & \textbf{14.77} & 16.22 & 16.14 & 19.66 & 19.42 \\ \hline
		\textbf{7.73} & 7.83 & 8.10 & 9.17 & 12.89 & 11.61 \\ \hline
		5.74 & \textbf{5.51} & 5.89 & 7.73 & 8.28 & 7.74 \\ \hline
		\textbf{4.13} & 5.97 & 5.41 & 6.00 & 12.83 & 10.64 \\ \hline
		\textbf{2.68} & 3.11 & 3.24 & 3.89 & 6.68 & 5.36 \\ \hline
		7.03 & \textbf{6.68 }& 7.13 & 6.98 & 17.10 & 12.64 \\ \hline
		\textbf{7.62} & 7.88 & 8.58 & 11.50 & 14.09 & 13.07 \\ \hline
		\textbf{11.15} & 11.69 & 12.02 & 11.80 & 15.57 & 15.04 \\ \hline
		\textbf{3.21} & 3.80 & 4.32 & 5.97 & 8.28 & 7.66 \\ \hline
		12.52 & \textbf{12.06} & 12.41 & 14.47 & 14.68 & 14.87 \\ \hline
		\textbf{	1.89} & 2.20 & 2.64 & 3.64 & 4.75 & 4.02 \\ \hline
		\textbf{2.20} & 2.46 & 2.66 & 4.56 & 5.10 & 4.68 \\ \hline
		\textbf{1.04} & 1.69 & 1.72 & 1.95 & 3.21 & 2.37 \\ \hline
		14.17 & \textbf{13.19} & 14.17 & 14.29 & 18.66 & 17.32 \\ \hline
		\textbf{5.77 }& 5.94 & 5.71 & 10.18 & 12.27 & 10.31 \\ \hline
		\textbf{3.48} & 4.15 & 4.73 & 5.30 & 8.99 & 8.28 \\ \hline
		\textbf{8.72} & 8.98 & 10.50 & 10.83 & 16.69 & 15.26 \\ \hline
		19.26 &\textbf{ 15.23} & 16.30 & 17.69 & 20.01 & 19.80 \\ \hline
		\textbf{5.02} & 5.21 & 5.29 & 5.94 & 11.18 & 8.33 \\ \hline
		\textbf{1.72} & 1.82 & 2.22 & 2.09 & 3.29 & 2.72 \\ \hline
		\textbf{5.52} & 6.61 & 6.59 & 9.75 & 13.63 & 11.06 \\ \hline
		\textbf{7.36} & 8.73 & 8.68 & 13.29 & 23.28 & 15.15 \\ \hline
		5.10 &\textbf{ 4.78} & 5.41 & 6.63 & 7.07 & 7.07 \\ \hline
		\textbf{2.55} & 3.16 & 3.34 & 4.67 & 7.14 & 5.39 \\ \hline\hline
		N.A. & 0.19 & 0.003 & $<10^{-4}$ & $<10^{-7}$ & $<10^{-7}$ \\ \hline
	\end{tabular}
\end{table}

\begin{figure}[h!]
	\centerline{\includegraphics[width=0.95\linewidth]{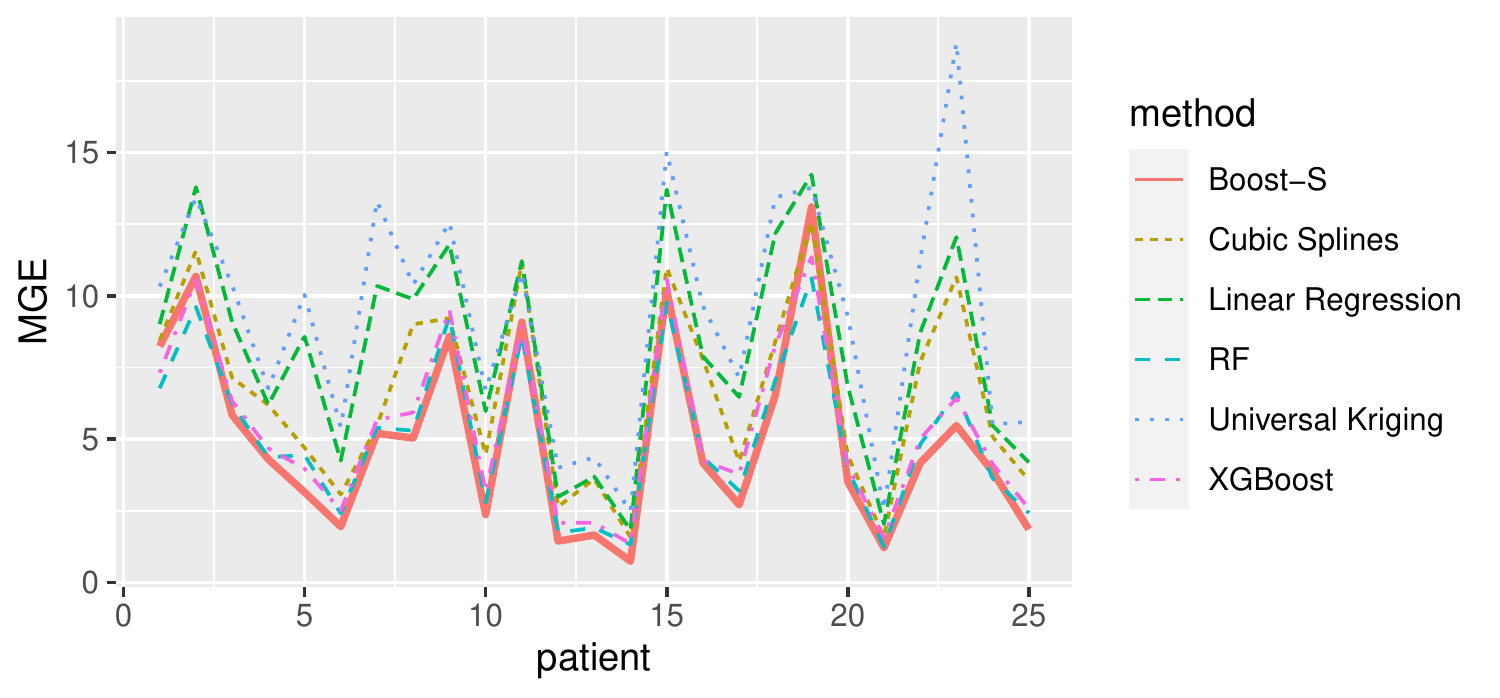}}
	\vspace{-8pt}
	\caption{Mean Gross Error (MGE) of the out-of-sample predictions for 25 patients using 6 different methods}
	\label{fig:mge}
\end{figure}

\begin{figure}[h!]
	\centerline{\includegraphics[width=0.95\linewidth]{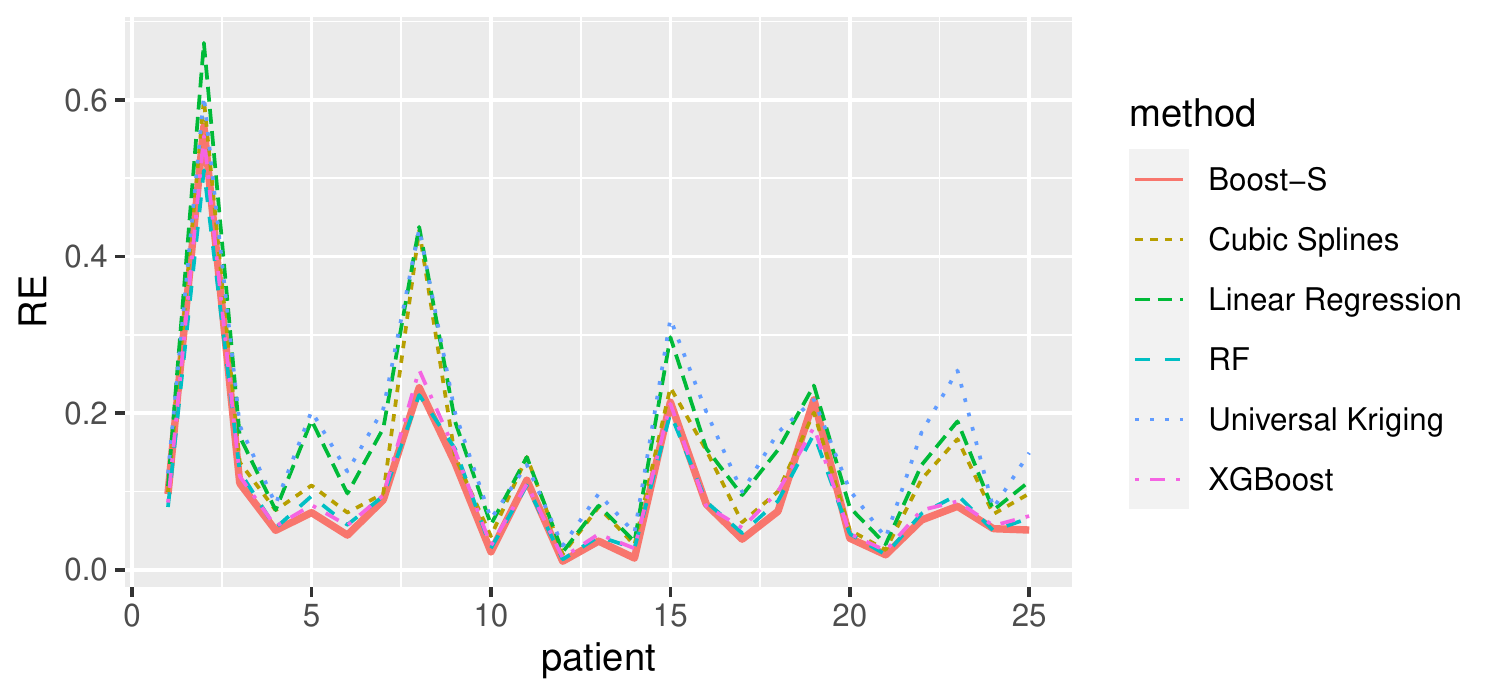}}
	\vspace{-8pt}
	\caption{Relative Error (RE) of the out-of-sample predictions for 25 patients using 6 different methods}
	\label{fig:re}
\end{figure}

\begin{figure}[h!]
	\centerline{\includegraphics[width=0.95\linewidth]{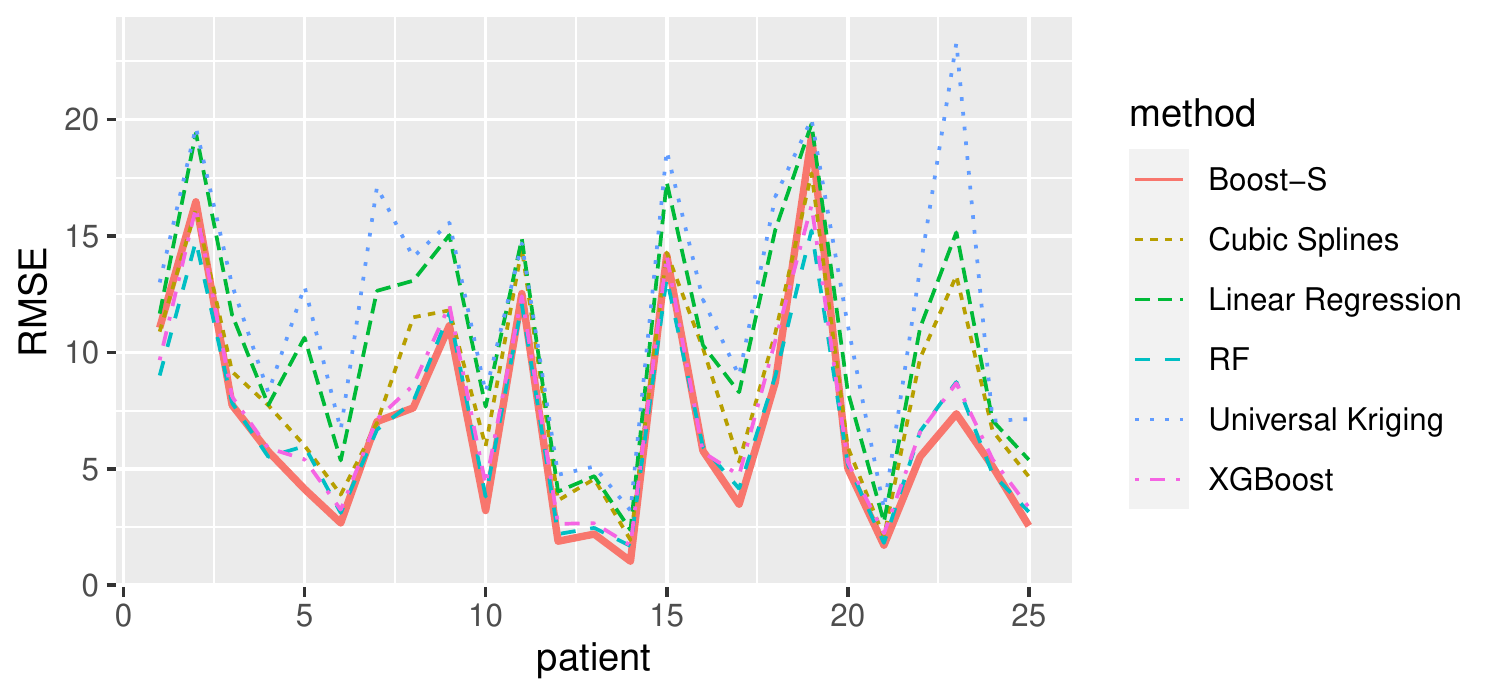}}
	\caption{Rooted Mean Squared Error (RMSE) of the out-of-sample predictions for 25 patients using 6 different methods}
	\label{fig:mse}
\end{figure}

 Figures \ref{fig:mse}, \ref{fig:mge} and \ref{fig:re} respectively show the MGE, RE and RMSE of the out-of-sample predictions for the 25 patients. Such visualizations provide a more holistic perspective on the performance of the six candidate methods.

\vspace{-8pt}

\section{Conclusion} \label{sec:conclusion}
This paper proposed a new gradient Boosted Trees algorithm for Spatial Data with covariate information (Boost-S). It has been shown that the Boost-S successfully integrates spatial correlation into the classical framework of gradient boosted trees. A computationally-efficient algorithm as well as the technical details have been presented. The Boost-S algorithm grows individual trees  by solving a regularized optimization problem, where the objective function involves two penalty terms on tree complexity and takes into account the underlying spatial correlation. The advantages of the proposed Boost-S, over five other commonly used approaches, have been demonstrated using real datasets involving the spatially-correlated FDG-PET imaging data collected during cancer chemoradiotherapy. 

\vspace{8pt}
\section{Acknowledgment}
This investigation was supported in part by National Institutes of Health grant R01CA204301.
%
%
%
%

%
\vspace{30pt}
\bibliographystyle{asa}
\bibliography{References}

\end{document}